\documentclass[journal]{IEEEtran}
\IEEEoverridecommandlockouts
\usepackage{cite}
\usepackage{amsmath,amssymb,bm}
\usepackage{graphicx}
\usepackage{textcomp}
\usepackage{xcolor}

\usepackage{hyperref}

\begin{document}

\title{A Unified Framework for Underwater Metaverse with Optical Perception}

\author{Jingyang Cao, Mu Zhou,~\IEEEmembership{Senior Member,~IEEE}~ Jiacheng Wang, Guangyuan Liu, Dusit Niyato,~\IEEEmembership{Fellow,~IEEE}, Shiwen Mao,~\IEEEmembership{Fellow,~IEEE}, Zhu Han,~\IEEEmembership{Fellow,~IEEE}, and Jiawen Kang

\thanks{J. Cao is with the School of Communications and Information Engineering, Chongqing University of Posts and Telecommunications, China and with the School of Computer Science and Engineering, Nanyang Technological University, Singapore (e-mail: N2308615A@e.ntu.edu.sg).  

M. Zhou is with the School of Communications and Information Engineering, Chongqing University of Posts and Telecommunications (e-mail: zhoumu@cqupt.edu.cn).

J. Wang, G. Liu, and D. Niyato are with the School of Computer Science and Engineering, Nanyang Technological University, Singapore (e-mail: jiacheng.wang@ntu.edu.sg, DNIYATO@ntu.edu.sg ).

S. Mao is with the Department of Electrical and Computer Engineering, Auburn University, Auburn, AL, 36849-5201 USA (e-mail: smao@ieee.org)

Zhu Han is with the Department of Electrical and Computer Engineering, University of Houston, Houston, TX 77004 USA (e-mail: hanzhu22@gmail.com)

J. Kang is with the School of Automation, Guangdong University of Technology, China (e-mail: kavinkang@gdut.edu.cn)
}

}

\maketitle

\begin{abstract}
With the advancement of AI technology and increasing attention to deep-sea exploration, the underwater Metaverse is gradually emerging. This paper explores the concept of underwater Metaverse, emerging virtual reality systems and services aimed at simulating and enhancing virtual experience of marine environments. First, we discuss potential applications of underwater Metaverse in underwater scientific research and marine conservation. Next, we present the architecture and supporting technologies of the underwater Metaverse, including high-resolution underwater imageing technologies and image processing technologies for rendering a realistic virtual world. Based on this, we present a use case for building a realistic underwater virtual world using underwater quantum imaging-generated artificial intelligence (QI-GAI) technology. The results demonstrate the effectiveness of the underwater Metaverse framework in simulating complex underwater environments, thus validating its potential in providing high-quality, interactive underwater virtual experiences. Finally, the paper examines the future development directions of underwater Metaverse, and provides new perspectives for marine science and conservation.
\end{abstract}


\section{Introduction}
\label{introduction}

The COVID-19 pandemic, along with the growth of virtual entertainment and quick progress in digital technology, has led to a profound transformation in societal interaction. This sparks considerable interest in digital experiences and greatly helps the developement of the Metaverse. The Metaverse, as a digitally blended space that merges the virtual and the real, has become a new platform for people to connect, communicate and entertain\cite{Lim2023}. In this wave, technology giants, such as Facebook, play a crucial role in Metaverse construction, striving to build digital cities, natural landscapes, and virtual social spaces.\footnote{\href{https://www.facebook.com/business/metaverse}{https://www.facebook.com/business/metaverse}} Apple Vision Pro seamlessly blends digital content with physical space.\footnote{\href{https://www.apple.com/apple-vision-pro/}{https://www.apple.com/apple-vision-pro/}} However, these efforts primarily focus on the development of the land Metaverse, overlooking another dimension of underwater Metaverse exploration, which is a novel concept focusing on creating and simulating underwater environments, encompassing shallow seas, deep oceans, lakes, glaciers, and more. Through highly realistic simulation technologies, it immerses users in a virtual underwater world, offering a completely new approach to explore and experience marine ecosystems.\footnote{\href{https://www.metaocean.space/}{https://www.metaocean.space//}}

The usefulness of the underwater Metaverse lies in its ability to create unique immersive and interactive underwater experiences beyond the limitations of physical reality. It enables people to explore, experience, and interact with the marine environment and its ecological content in unprecedented ways. In this virtual underwater world, users can embark on deep-sea adventures, observe rare marine lives, and even conduct scientific research, with rich experiences that may be difficult to obtain in the physical world.\footnote{\href{https://www.yaswaterworld.com/en/experiences/underwater-vr}{https://www.yaswaterworld.com/en/experiences/underwater-vr}} Furthermore, there are many applications of the underwater Metaverse, such as virtual restoration  of underwater cultural heritage, virtual design for underwater resource exploration, simulation of underwater disasters and rare deep-sea creatures. Additionally, the underwater Metaverse fosters new forms of social interaction, allowing people to collaborate in this virtual seabed world. Its advantages are mainly reflected in the following aspects:

\begin{itemize}
\item  {{\em Overcoming Physical Limitations}: Real underwater exploration faces severe challenges from natural conditions such as extreme depth, pressure, temperature and light, which are significant limitations for both humans and equipment\cite{Liu2020}. The Metaverse provides a safe, risk-free virtual environment, allowing users to explore the deep-sea without constraints of the physical world.
}
\end{itemize}

\begin{itemize}
\item  {{\em Educational and Research Tool}: The underwater Metaverse can be a powerful tool for education and research. In the form of digital twin, it can accurately simulate underwater environments and physical conditions, which are crucial for scientific experiments and other fields.\footnote{\href{https://digitaltwinocean.mercator-ocean.eu/}{https://digitaltwinocean.mercator-ocean.eu/}} Maritime education can leverage this platform to learn and study various marine phenomena and biological behaviors.
}
\end{itemize}

\begin{figure*}[ht]
  \centering
  \includegraphics[width=0.75\linewidth]{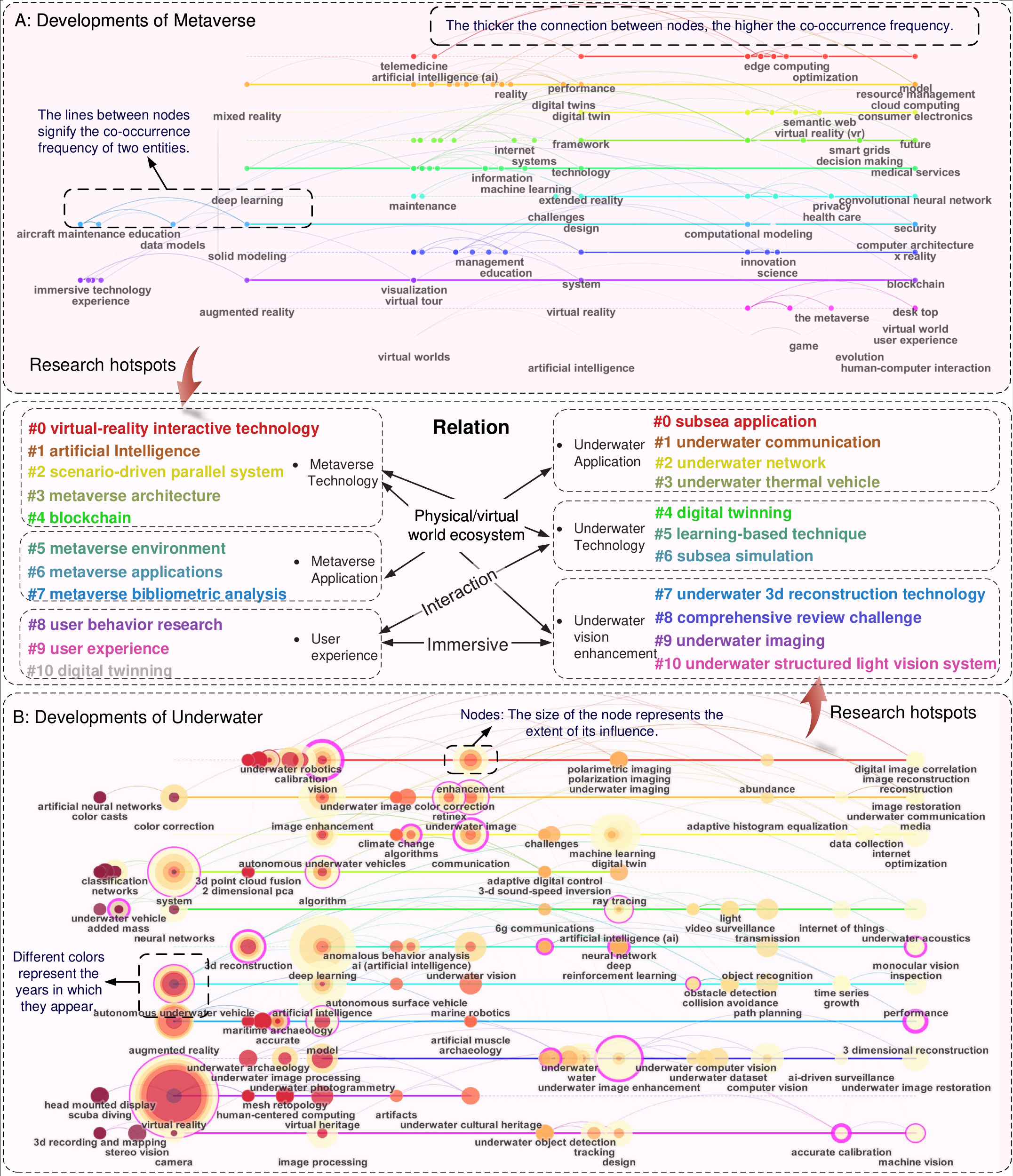}
  \caption{Keywords clustering results and timeline distribution in underwater research and Metaverse studies. Keyword clustering analysis highlights the relationship between research directions and key terms. As shown in Fig. 1(A), in Metaverse research, virtual reality interaction technology forms the largest cluster and has an early stage in the field. The rise of blockchain and artificial intelligence technologies have propelled the development of the Metaverse's structure and environment studies. Increasingly, there is a focus on user experience, emphasizing enhanced interactivity, realism, and engagement to create more immersive and captivating virtual environments. As shown in Fig. 1(B), in underwater research, the clustering analysis identified 10 clusters, with underwater applications being the largest and earliest-starting cluster. With the development of underwater networks and detectors, there is growing focus on constructing reliable systems for underwater data collection and analysis. Consequently, underwater image reconstruction, digital twin, and visual analysis have become the main research subjects (Data from Web of Science: https://www.webofscience.com).}
  \label{fig:1}
\end{figure*}

\begin{itemize}
\item  {{\em Unlimited Exploration Range}: The underwater Metaverse can create highly realistic underwater environments, giving users a strong sense of immersion, especially as they can explore areas that are difficult and impossible to reach in the real world, such as deep-sea trenches and remote seabed volcanoes.\footnote{\href{https://www.marum.de/en/Discover/Virtual-Deep-Sea.html}{https://www.marum.de/en/Discover/Virtual-Deep-Sea.html}} As human exploration of oceans continues and interest in marine ecology grows, the underwater Metaverse is becoming an important bridge connecting reality and virtuality, nature and technology, offering endless possibilities for future ocean exploration, military simulation, and environmental protection.
}
\end{itemize}

However, we are still far from realizing the underwater Metaverse vision. One characteristic of underwater environments is the significant reduction of light, especially in deep-sea areas. Low light conditions pose higher demands on the imaging technology. This requires the underwater Metaverse  to accurately capture and reconstruct target images under harsh conditions, including sensitive marine life and complex seabed terrain. However, existing underwater imaging technologies are limited in capturing such details, particularly in maintaining high resolution and clarity in low-light conditions \cite{Li2023}. Additionally, the dynamic nature of underwater environments adds to the challenge. Marine ecosystems are constantly changing, with multiple factors such as water currents and biological activity, which requires the underwater Metaverse to create a dynamic and highly interactive virtual environment. Finally, with increasing attention to data privacy and security, another significant challenge that we face is how to provide a rich and engaging underwater Metaverse experience while protecting user privacy. Facing these challenges, we conduct literature analysis to delve into the key areas of underwater and Metaverse research, revealing the technological advancements and potential directions in addressing these difficulties.  Fig.~\ref{fig:1} illustrates the emerging research topics related to underwater and the Metaverse. Underwater research focuses on image reconstruction, digital twin and vision, which are crucial for accurate marine simulation and analysis. In Metaverse research, areas including artificial intelligence, user experience, and environmental simulation lay the foundation for an interactive, intelligent virtual underwater environment. 

Overall, the contributions of this paper are as follows:

\begin{itemize}
\item  {We propose the concept of underwater Metavers. Underwater Metaverse is a virtual underwater world, it includes multiple aspects such as visual, perception, data collection, and interaction in the underwater environments. Then, we propose and design the first underwater Metaverse architecture. This architecture focuses on several key factors such as engine and safety, offering feasibility and operability for the comprehensive implementation of the underwater Metaverse.
}
\end{itemize}

\begin{itemize}
\item  {We propose the quantum imaging-generative artificial intelligence (QI-GAI) technology. This technology uses quantum imaging to achieve high-precision and high-fidelity capture of underwater objects under harsh conditions, overcoming the limitations of existing imaging technologies in underwater environments. On this basis, GAI is then used to simulate and generate a highly realistic underwater environment. Through a case study, we verify the feasibility of the proposed technology and provide strong support for future applications of the underwater Metaverse.
}
\end{itemize}

\begin{itemize}
\item  {We discuss some unresolved issues. The exploration of these issues provides future research directions for academia and industry, to promote in-depth development in the field of underwater Metaverse.
}
\end{itemize}

\section{The Underwater Metaverse: Defination, Applications, and Architecture}
\label{GAI-and-quantum-imaging-technologies}

\subsection{Definition}
\label{definition-and-architecture}

The underwater Metaverse defined as {\textit{a submerged extension of the digital universe, which intricately weaves together complex and dynamic underwater environments. It encompasses immersive, interaction, and boundless aquatic virtual environments, explored through user-directed avatars in a deep-sea setting.}} In the following, we explain each feature in this definition. 

\begin{itemize}
\item   {{\textit{Immersive:}} An immersive experience in the underwater Metaverse makes users feel fully absorbed in a virtual environment, achieved through high-quality imagery and simulations of physical characteristics, creating a sensation of truly exploring ocean environments.}
\end{itemize}

\begin{figure*}[ht]
  \centering
  \includegraphics[width=0.65\linewidth]{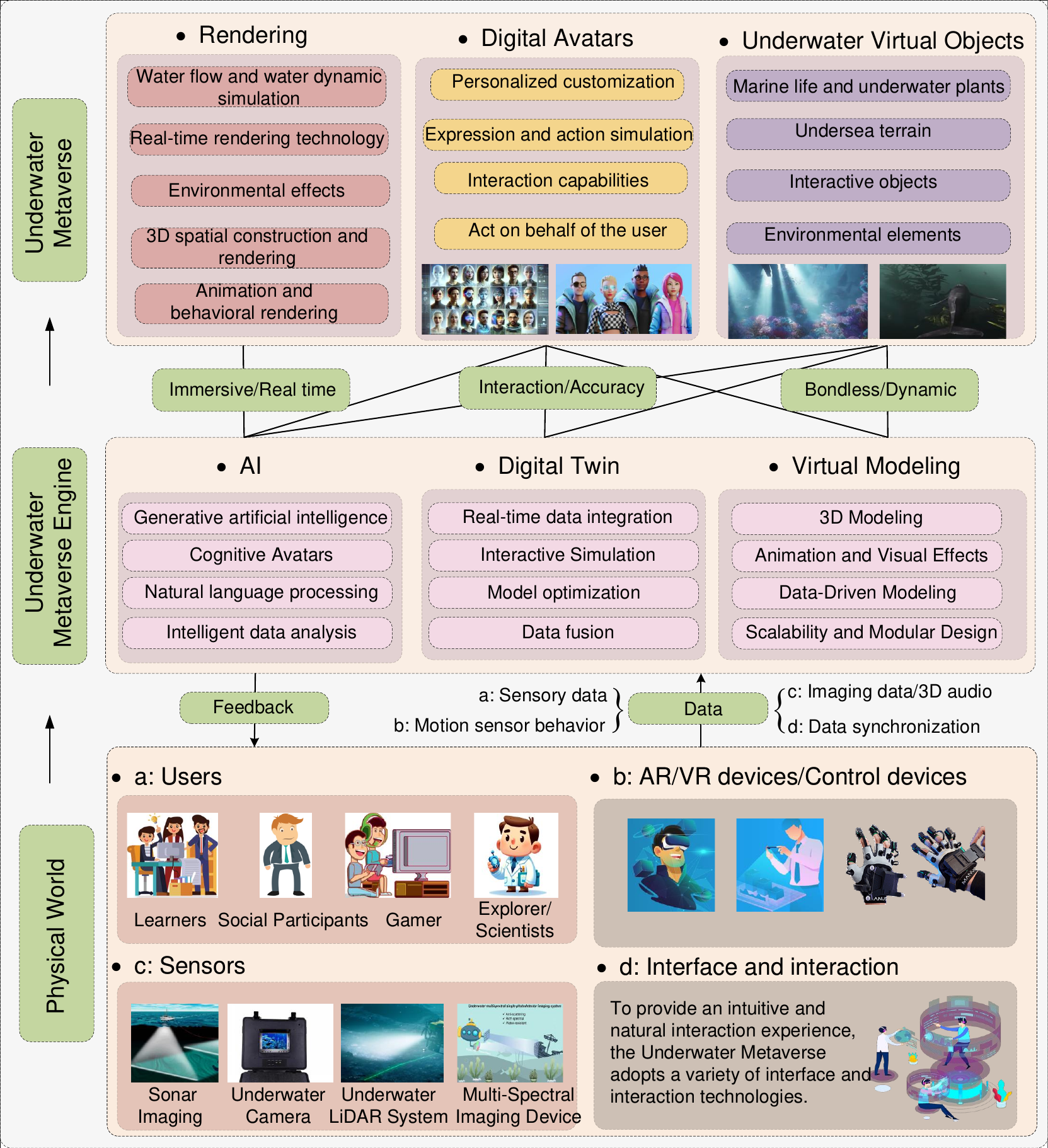}
  \caption{We can divide underwater Metaverse framework into three key layers: the Physical World, the Underwater Metaverse Engine, and the Virtual World. Each of these layers not only carries out its unique functions and responsibilities but also closely interconnects with each other, together forming a complete ecosystem of the underwater Metaverse. The precision and efficiency of the physical layer are crucial for constructing a virtual environment with a strong sense of realism. The engine layer not only ensures the dynamics and authenticity of the virtual world but also provides the necessary computational support to facilitate complex virtual interactions. The collaborative effort of rendering technology, digital avatars, and virtual objects creates a richly diverse and highly interactive underwater world}
  \label{fig:2}
\end{figure*}

\begin{itemize}
\item  {{\textit{Interaction:}} Interaction allows users to engage with virtual marine life, explore underwater realms, and collaborate with each other, enhancing engagement and making the experience more lively and realistic.}
\end{itemize}

\begin{itemize}
\item  {{\textit{Dynamic:}}
Dynamic refers to the continuous changing and active characteristics of the virtual environment. Users can observe and experience varying surroundings and conditions virtually.}
\end{itemize}

\begin{itemize}
\item  {{\textit{Boundless:}} Boundless means exploring without the limitations of geography, safety, and other real-world constraints, allowing users to freely navigate through an unrestricted virtual marine landscape, from shallow waters to deep seas, and even to places unreachable in reality.}
\end{itemize}

\subsection{Applications of the Underwater Metaverse}
\label{applications}

Diving into the underwater Metaverse, we uncover four fascinating applications that are revolutionizing our interaction with the marine world.

\begin{itemize}
\item {\textit{Virtual Restoration and Exhibition of Underwater Cultural Heritage:}} The underwater Metaverse offers a unique way to explore and protect submerged cultural heritage, such as ancient ships, sunken ships and submerged city sites.\footnote{\href{https://www.euronews.com/green/2023/07/25/photogrammetry-virtual-reality-and-ecotourism-exploring-and-preserving-our-underwater-heri}{https://www.euronews.com/green/2023/07/25/photogrammetry-virtual-reality-and-ecotourism-exploring-and-preserving-our-underwater-heri}} Utilizing high-precision imaging and virtual reality technologies, these relics can be accurately restored and presented to the public. This approach not only makes remote and deep-sea sites more accessible but also provides a risk-free environment for archaeological research. Through the underwater Metaverse, we can recreate history, serving as a powerful tool for education and cultural heritage restoration.
\end{itemize}

\begin{itemize}
\item {\textit{Virtual Planning and Design for Underwater Engineering and Resource Exploration:}} The underwater Metaverse offers a robust tool for planning and designing underwater engineering projects. Within this virtual environment, engineers can simulate the construction process of seabed pipeline laying, deep-sea mining, and other engineering works. This method allows for the early detection of potential design issues and testing the performance of equipment and structures under extreme deep-sea conditions. Additionally, resource exploration in the underwater Metaverse, such as deep-sea mineral resources, can assess potential resource locations and mining value without disturbing the marine environment, providing key information for sustainable resource utilization.
\end{itemize}

\begin{itemize}
\item  {\textit{Underwater Disaster Simulation and Emergency Response Training: }}The underwater Metaverse can be used to simulate marine disasters, such as tsunamis, seabed earthquakes, and oil and gas leaks. In these virtual environments, rescue teams can conduct emergency response training, study the potential impact of disasters, and develop effective response strategies. For example, by simulating oil and gas leak scenarios, different containment and cleanup strategies can be tested, thus enhancing the efficiency and safety of real-world disaster recovery.
\end{itemize}

\begin{itemize}
\item  {\textit{Deep-Sea Rare Creatures Simulation:}} The underwater Metaverse, with its three-dimensional modeling technology, can create realistic deep-sea environments, including complex terrains like deep-sea trenches and hydrothermal vents. In these simulated environments, rare deep-sea creatures such as deep-sea octopuses, unique fish species, and other not fully studied organisms can be virtually recreated. These simulations not only replicate the appearance of these creatures but also include their behavioral patterns, habits, and interactions with the environment.
\end{itemize}

\subsection{Underwater Metaverse Architecture}
\label{definition-and-architecture}

The architecture of the underwater Metaverse is presented in Fig.~\ref{fig:2}
\subsubsection{\textbf{Physical-Virtual World Interaction}}

The relationship between these two layers is iterative and interactive. As users engage with the virtual layer, their actions can trigger changes that influence data collection and analysis in the physical layer, creating a feedback loop. The integration ensures that virtual representation of underwater objects and animals remains accurate and up-to-date to users exploring the dynamic ocean environment. It includes:

\begin{itemize}
\item  {{\textit{Data Collection and Transmission:}} The physical layer forms the foundation of the entire underwater virtual universe system. Various sensors and devices, such as underwater cameras, sonar, and other measurement tools, are responsible for collecting environmental data. The collected data is then transmitted to the engine layer. In addition, underwater communications are crucial for data transmission, including acoustic communication, optical communication, and radio frequency communication. The engine layer, serving as the center for data processing and analysis, transforms these physical data into information usable in the virtual world.}
\end{itemize}

\begin{itemize}
\item  {{\textit{User Interaction and Feedback: }}Users interact in the virtual world through input devices such as VR, controllers, and haptic devices \cite{Xu2023}. These devices enable users to dive, explore, and interact with the virtual underwater environment and other users. Actions of users within the virtual underwater world are captured and processed by the engine layer, which can influence the output of the engine layer, in turn, affect the operation of devices in the physical world.}
\end{itemize}

\begin{itemize}
\item  {{\textit{Two-Way Dynamic Interaction:}} User activities in the virtual world not only affect the engine layer but can also be fed back to the physical world, creating a two-way interaction mechanism. For example, user exploration behaviors in the virtual underwater world can guide the path of submarines in the ocean. Similarly, changes in the physical world (e.g., tides and current) are instantly reflected in the virtual underwater world.}
\end{itemize}

\begin{itemize}
\item  {{\textit{Augmented Reality and Integrated Experience: }}Elements of the physical and virtual worlds can be combined in an augmented reality environment, providing users with a richer and more intuitive interactive experience\cite{Lee2021}. Users can interact with virtual objects in a real environment, making underwater exploration and research more vivid and interactive.}
\end{itemize}

\subsubsection{\textbf{Underwater Metaverse Engine}}
The engine layer is the core processing and analytical layer within the framework of the underwater Metaverse, acting as a bridge between the physical world and the virtual world. The primary task of this layer is to receive data from the physical world layer, conduct in-depth analysis and processing, and based on this, construct and maintain a dynamic underwater Metaverse virtual environment. Specifically, it includes:

\begin{itemize}
\item  {{\textit{Artificial Intelligence:}} AI technology is used to process and analyze large amounts of data collected from the underwater environment, including marine biological activities and hydrogeological data. AI technology can identify patterns, predict environmental changes, and automate decision-making processes \cite{JWang2023}. This technology helps optimize the construction and management of the underwater Metaverse environment, improving the system's responsiveness and level of intelligence.}
\end{itemize}

\begin{figure*}[ht]
  \centering
  \includegraphics[width=\linewidth]{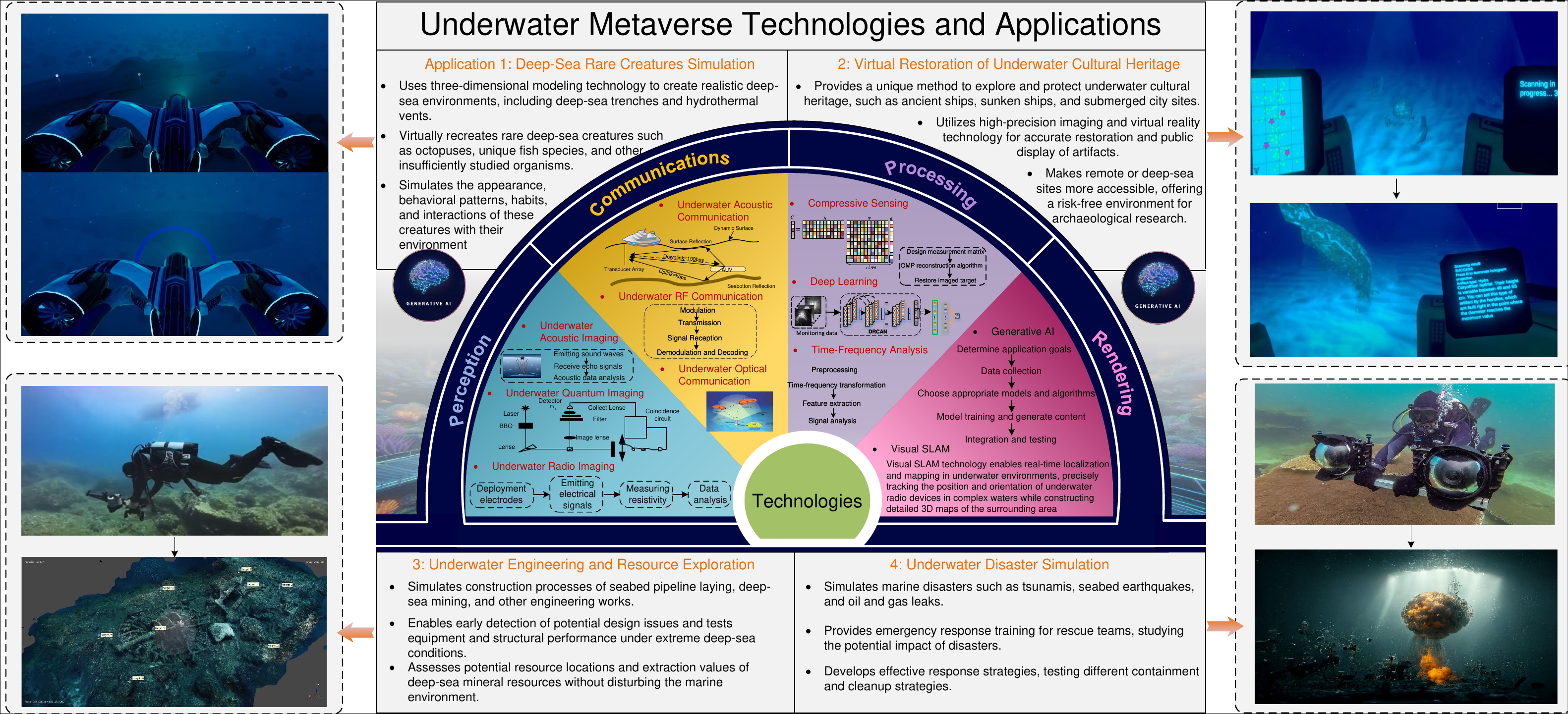}
  \caption{Underwater Metaverse technology and applications represent a revolutionary advancement, combining key Metaverse technologies with ocean exploration to offer an unprecedented way to understand and experience the underwater world\cite{Lee2021}. This field integrates various technologies such as VR, AR, visual, AI, and generative algorithms to create a realistic and highly interactive virtual underwater environment. With the continuous advancement of technology and the growth of applications, the underwater Metaverse is expected to play an increasingly important role in the future.}
  \label{fig:3}
\end{figure*}

\begin{itemize}
\item  {{\textit{Digital Twin:}} Digital twin technology in the engine layer of the underwater Metaverse is used to create an accurate digital replica of the physical underwater environment.\footnote{\href{https://ocean-twin.eu/digital-twins}{https://ocean-twin.eu/digital-twins}} This includes virtual models of marine terrain, underwater structures, and marine life. With real-time and periodic updates, these digital twin models can reflect the current state of the real environment, providing users with real-time environmental information and a rich interactive experience.}
\end{itemize}

\begin{itemize}
\item  {{\textit{Virtual Modeling:}} Virtual modeling is used to create and maintain various elements in the virtual environment. This includes virtual marine creatures, simulated water flow and lighting effects, and shallow and deep-water environments. Advanced modeling technology ensures that the virtual environment is not only visually realistic but also capable of interacting in real-time with user's actions and choices.}
\end{itemize}

\section{Enabling Underwater Metaverse Technologies}
\label{GAI-and-quantum-imaging-technologies}
The underwater Metaverse includes four key technologies: perception, communications, processing and rendering, as shown in the Fig.~\ref{fig:3}

%\subsection{Key technologies}
%\label{Key technologies}

%\subsubsection{\textbf{Perception}}

\subsection{Perception}

 In the underwater Metaverse, perception is primarily used for capturing and interpreting environmental information. The imaging system of visual perception is particularly critical as it directly affects our cognition and understanding of the underwater world. It can be divided into three main parts.

\subsubsection{\em {Underwater Acoustic Imaging}}
Acoustic imaging is a method of detection and imaging using sound waves. This technology is particularly effective in underwater environments, as sound waves can travel long distances in water, detecting farther objects and terrains \cite{Sutton2023}. Sonar technology works by emitting sound waves and capturing their reflections, and analyzing the reflections to determine the location, shape, and other characteristics of objects. The resolution of underwater acoustic imaging depends on the frequency of the sound waves used. Higher frequencies provide finer details, but at the cost of reduced detection range. This technology has wide applications in underwater navigation, terrain mapping, and detection of underwater life.

\subsubsection{\em {Underwater Radio Imaging}}
Underwater radio imaging technology utilizes radio waves to detect and image underwater environments. This technology primarily relies on the interaction between radio waves and underwater objects \cite{Smolyaninov2023}, and its resolution is limited by the propagation characteristics of radio waves in water. When radio waves encounter underwater objects, their propagation characteristics change, such as reflection, refraction and scattering, thereby creating specific signal patterns at the receiver. Underwater radio imaging systems identify and image underwater objects by analyzing these altered signals. This method is highly effective for detecting the structure and characteristics of underwater objects, such as their shape, size, and material, making it highly valuable for applications in underwater navigation, communication, and other related fields. In addition, this technology is suitable for shallow water or near-surface applications.

\subsubsection{\em {Underwater Quantum Imaging}}
Due to the absorption and scattering of light by water, its propagation in water is severely affected, requiring light imaging technology to adapt to this special environment. QI technology plays a significant role in this aspect \cite{Cao2022}. It utilizes the time-space correlation properties of entangled photon pairs to enhance the accuracy and efficiency of imaging. Compared with traditional underwater optical imaging, this technology shows significant advantages in reducing noise and improving image clarity. These quantum properties allow light imaging systems to capture high-quality images even in very low light conditions, thus visually representing the underwater world more accurately.

\subsection{Communications}
In the construction of the underwater Metaverse, underwater communication technology is the infrastructure for achieving interaction and immersive experiences.

\subsubsection{\em {Acoustic-based Underwater Communication}}
Underwater acoustic communication utilizes the propagation characteristics of sound waves in water for data transmission. This technology\footnote{\href{https://acomms.whoi.edu/}{https://acomms.whoi.edu/}} is primarily used for long-distance communication and data transfer in the underwater Metaverse, especially suitable for extensive underwater environmental exploration. Information is transmitted using modulation techniques such as Frequency Shift Keying (FSK), Phase Shift Keying (PSK), and Orthogonal Frequency-Division Multiplexing (OFDM). These techniques encode information by altering the frequency, amplitude, and phase of sound waves. Taking into account the impact of temperature, salinity and pressure variations on sound wave propagation in underwater environments, the communication system uses sound velocity profiling and multipath effect correction models to address unique underwater challenges such as multipath effects and velocity profile changes, thereby improving the accuracy and reliability of communication.

\subsubsection{\em {RF-based Underwater Communication}}
Despite the significant limitations of underwater environments on RF signals, under certain conditions, such as using low-frequency bands, RF communication can still be effective over short distances. Key technologies\cite{P2021} include short-range radio wave propagation models, frequency selection, and modulation techniques such as Amplitude Modulation (AM) and Frequency Modulation (FM). Considering the attenuation characteristics of radio waves in water, the system comprises efficient signal encoding and decoding techniques, along with robust signal processing capabilities to overcome high attenuation and noise in underwater environments. RF communication is particularly effective in shallow water and near-surface applications, thus the algorithms also involve optimizing transmission distance and power to achieve optimal underwater communication performance.

\subsubsection{\em {Optical-based Underwater Communication}}
Due to its high-speed and low-latency characteristics, underwater optical communication technology\footnote{\href{https://www.shimadzu.com/underwater/index.html}{https://www.shimadzu.com/underwater/index.html}} offers significant advantages in transmitting high-quality video and image data. This technology employs laser modulation techniques such as Pulse Position Modulation (PPM), Pulse Width Modulation (PWM), and signal enhancement techniques for optical receivers. Optical communication system also includes beamforming and directional control techniques to minimize propagation loss, due to the scattering and absorption of light waves in water. In the underwater Metaverse, this communication technology is used for fast and efficient data exchange, providing a more realistic and enriched visual experience.

%\subsubsection{\textbf{Processing}}
\subsection{Processing}
In the underwater Metaverse, it is necessary to analyze and process data collected from perception technologies such as acoustic imaging, radio imaging, and quantum imaging. The purpose of this process is to extract useful information from raw data to better understand and utilize the characteristics of the underwater environment. In underwater settings, due to factors like light attenuation, sound wave scattering, and electrical signal interference, the processing of imaging data is particularly complex. Here are several primary data processing techniques:

\subsubsection{\em {Deep Learning in Underwater Acoustic Imaging}}
In underwater acoustic imaging, sound waves are used to detect underwater environments, and deep learning can effectively process the acoustic data\cite{Kul2022}. First, the underwater acoustic data needs to be preprocessed, which includes noise filtering, contrast enhancement, and image normalization, to improve the quality of the original underwater sonar images and make them more suitable for training and prediction with deep learning models. Second, feature extraction is performed. For example, Convolutional Neural Networks (CNNs) can automatically extract and learn hierarchical features of images. Then, model selection and training are conducted. For image classification, architectures such as ResNet and VGG can be used. For object detection tasks, models like Faster R-CNN, and Single Shot MultiBox Detector (SSD) are employed. Lastly, post-processing is performed, which may include filtering, non-maximum suppression and threshold setting, to enhance the accuracy and readability of the final results.

\subsubsection{\em {Time-Frequency Analysis in Underwater Radio Imaging}}
Time-Frequency Analysis technology\cite{Chen2002} can effectively decompose underwater radio signals and extract frequency components at different time instances by combining time domain and frequency domain analysis. In the preprocessing stage, frequency domain filtering can be employed, particularly utilizing band-pass filters to eliminate high and low-frequency noises in the signal. For the time-frequency analysis part, the sophisticated Continuous Wavelet Transform (CWT) could be utilized, which is suitable for processing non-stationary and time-variant signals. Then, the Hilbert-Huang Transform (HHT) shall be employed to further improve the precision of time-frequency analysis. Statistical methods can be used to extract key signal features, such as the energy distribution and entropy of the signal, and Spectral Peak Tracking (SPT) technology can be used to identify the main frequency components and their trends. In the post-processing stage, the extracted features are smoothed to eliminate noise interference, and data visualization techniques, such as time-frequency plots and spectrograms, are used to intuitively present the time-frequency characteristics of the signal. The advantage of this technology lies in its enhanced signal resolution, especially for non-stationary and rapidly changing underwater signals, which can more accurately distinguish and identify various types of underwater objects and structures.

\subsubsection{\em {Compressive Sensing in Underwater Quantum Imaging}}
In underwater QI, the image data typically produced are sparse in certain transform areas, meaning that the majority of coefficients are close to or equal to zero. In this process, it is necessary to design a measurement matrix, such as a Gaussian random matrix and Bernoulli random matrix, to perform underwater data acquisition. Signal reconstruction is the core of compressed sensing\cite{Hu2002}, we can utilize optimization algorithms such as Matching Pursuit (MP), Orthogonal Matching Pursuit (OMP), and Basis Pursuit (BP) to recover the original signal and image from a limited number of measurements. The goal is to minimize the sparsity of the reconstructed image while maintaining consistency with the observed data. Finally, the reconstructed data can be post-processed as needed, such as denoising and filtering, to enhance the imaging quality and clarity. The advantage of this method lies in its ability to efficiently reconstruct high-quality images from fewer data measurements, which is particularly crucial for underwater environments where data transmission and storage resources are limited, significantly reducing the dependency on expensive and complex sensors.

%\subsubsection{\textbf{Rendering}}
\subsection{Rendering}
Rendering technology can transform processed data into intuitive images and scenes, thereby providing users with highly realistic virtual environments. In the underwater Metaverse, rendering technology faces unique challenges, such as the need to accurately simulate underwater optical properties, dynamic environmental changes, and complex biological ecosystems. Here are several main rendering techniques applied in the underwater Metaverse:

\subsubsection{\em {Generative AI Driven Underwater Metaverse}}
Underwater images often suffer from quality issues like blurriness and color deviation due to factors such as light attenuation and water scattering. Generative AI technologies, including Generative Adversarial Networks (GANs) and Variational Autoencoders (VAEs) are employed to reconstruct high-resolution images from sparse data, effectively capturing complex details of underwater environments, such as terrain and biological entities.

\subsubsection{\em {Visual SLAM Driven Underwater Metaverse}}
Visual SLAM technology facilitates real-time localization and mapping in underwater environments, accurately tracking the position and orientation of devices while constructing detailed 3D maps.\footnote{\href{https://www.maxstblog.com/post/types-of-slam-and-application-ex}{https://www.maxstblog.com/post/types-of-slam-and-application-ex}} This technology enhances the spatial accuracy of underwater imaging data, providing a precise spatial reference. The combination of processed imaging data with the maps generated by visual SLAM allows for the creation of realistic underwater virtual environments. The precise localization capabilities of visual SLAM, along with efficient data processing, enable the development of an accurate and real-time underwater metaverse.

\section{Case Study}
\label{Case-study}

\subsection{Experimental Configuration}
\label{experimental-configuration}

\begin{figure*}[ht]
  \centering
  \includegraphics[width=\linewidth]{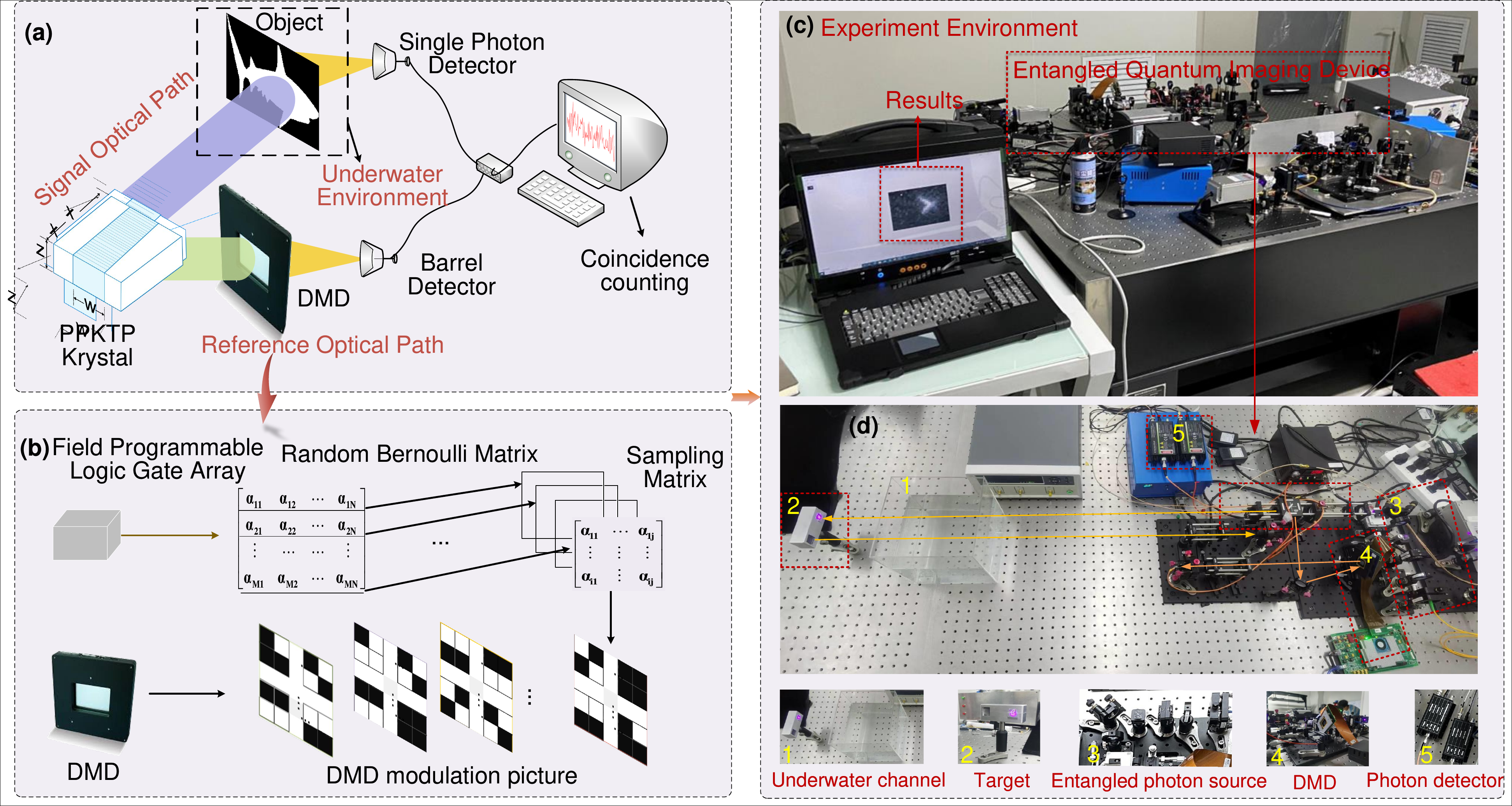}
  \caption{(a) is a schematic diagram of underwater entangled photon quantum imaging. (b) shows the loaded modulation pattern in DMD. In the quantum image, the white squares indicate areas on the DMD that remain open. Only reference photons in the open areas can be scanned by the DMD and collected by the single-photon detector. (c) and (d) represent the constructed imaging optical path and the underwater quantum imaging experimental setup.}
 \label{fig:4}
\end{figure*}

Due to light attenuation and interference from water impuriy, traditional imaging technology is difficult to obtain high-quality images for rendering underwater Metaverse targets, which has a significant impact on the visual experience of users. Therefore, we utilize the underwater QI technology, as shown in Fig.~\ref{fig:4}(a), which leverages entanglement, non-locality, and spatio-temporal correlations to achieve high-resolution imaging under low-light conditions, while also reducing the effects of water scattering and absorption. This technology not only improves the quality of underwater imaging but also captures more details in complex underwater environments, such as the deep-sea and turbid waters.

To validate the underwater quantum imaging, we conduct tests in an experimental environment. We set up an entangled QI platform based on compressed sensing and simulate the underwater environment, as shown in Fig.~\ref{fig:4}(c). This platform includes an entangled light source, signal and reference optical paths, a water tank, a Digital Micromirror Device (DMD), and two single-photon detectors for receiving entangled photons. In particular, we set up a water tank on the signal light path, so that the signal light can vertically pass through the water tank and directly illuminate the target object \cite{Cao2022}, as shown in Fig.~\ref{fig:4}(d). We simulate deep and shallow marine environments, thereby more accurately testing the effectiveness of our proposed imaging technology. Due to the equipment limitations, we use cards to simulate the targets. Then we reconstruct the image by obtaining coincidence count values of the real target. Through this method, the number of required measurements can be significantly reduced while maintaining high-quality image reconstruction.

\subsection{Performance Analysis}
\label{Performance-analysis}

\begin{figure*}[ht]
  \centering
  \includegraphics[width=\linewidth]{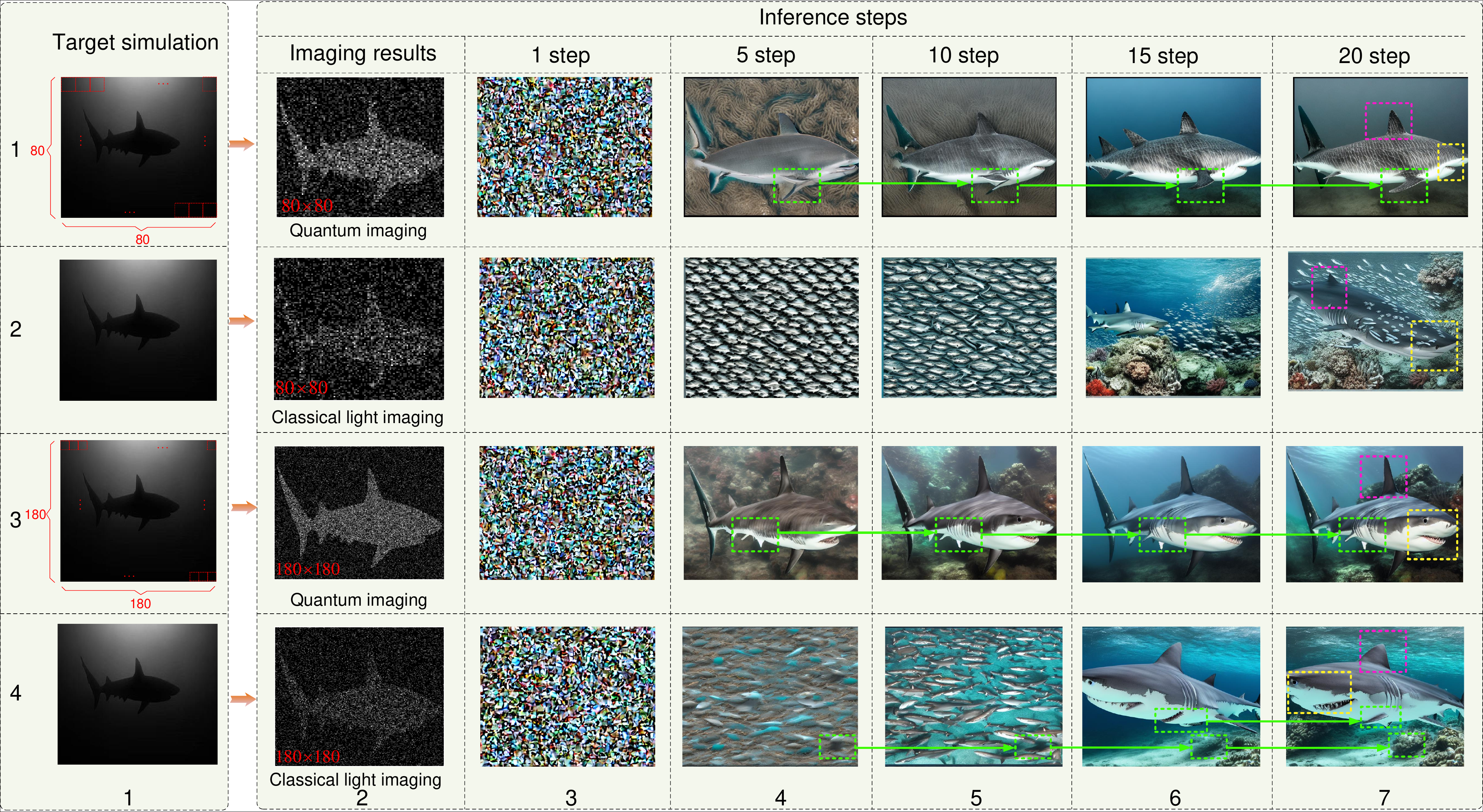}
  \caption{The impact of GAI steps on generated content. The images on the left are the input underwater quantum light source and classical light source images, and the image on the right is the corresponding generated content. As can be seen, more steps trigger more natural content, showing fuller details and contours, as shown in the green box.}
  \label{fig:5}
\end{figure*}

We then analyze image generation, using the virtual avatar of a shark in the underwater Metaverse as an example, to demonstrate how QI-GAI specifically supports the underwater Metaverse. The generated results are displayed in Fig.~\ref{fig:5}. We use the ControlNet model\footnote{\href{https://github.com/lllyasviel/ControlNet}{https://github.com/lllyasviel/ControlNet}} and the steps are set to 20. From the results, we can see that the diffusion model (GAI) can effectively generate virtual avatars based on underwater target data and instruction sets. With the increase in inference steps, the image quality can be significantly improved. Specifically, when only one inference step is executed, the generated content is merely noise, as shown by the result corresponding to step 1 in Fig.~\ref{fig:5}. As the number of inference steps increases, the shape begins to gradually emerge, and its details correspond to the real target. For example, in the results of the first row, as the steps increase, the details of the object gradually are enriched, and the background details are also gradually refined. Therefore, the model can generate images based on quantum images and user instructions, confirming the effectiveness of the QI-GAI technology, which can provide support for the underwater Metaverse.

Moreover, we compare images generated using underwater quantum light sources and classical light sources.  We set two groups of DMD with different bundled pixel values, $80\times 80$ and $180\times 180$, respectively. From the results in the second column, we can see that the resolution of underwater quantum imaging is higher than that of classical light sources, showcasing better underwater imaging results, and higher bundled pixel images have higher resolution. However, if the bundled pixel value is set too high, it can lead to lower image resolution due to the introduction of more stray photons. From the avatar images generated in the seventh column, we can see that QI-GAI generates better shark outlines and richer details, closer to the real target, as shown by the details in the yellow frame. From the results in the first and third rows, we can see that higher bundled pixel values display more shark details, such as the shark fins shown in the red frame. However, the generated shark forms vary, which can be guided by other parameters, e.g., environment conditions and user preferences. Moreover, multimodality techniques can be adopted for adjusting the variation.

\section{Future Directions}
\label{Open-issues-and-future-direction}

\subsection{Three-Dimensional Underwater Quantum Imaging}
Due to the superior penetration capabilities of quantum light sources, QI is highly effective in deep-water and low-visibility environments, which is of significant importance for the exploration and research of the underwater metaverse. However, the existing QI technology is limited to two dimensions. Therefore, in the construction of the underwater Metaverse, underwater three-dimensional QI technology is not only the key to achieving advanced visual effects but also central to enhancing user experience. Technologically, three-dimensional QI, utilizing the variance in flight times, can construct the z-axis information, i.e., using the multidimensional characteristics of quantum to increase the density of imaging information. Moreover, this imaging technology requires overcoming the existing optical limitations, developing quantum sensors and data processing algorithms that can operate stably under extreme underwater conditions, and optimizing the extraction and analysis of quantum information to ensure the accuracy and real-time updating of three-dimensional models.

\subsection{Balance of Computing Resources}
In future development of the underwater Metaverse, rational allocation and optimization of computational resources are particularly crucial. Since the underwater Metaverse involves real-time processing of massive data and complex simulation modeling\cite{Wang2023}, it requires effective distribution of computational tasks among various system nodes—including edge devices, underwater servers, and cloud infrastructure. Such allocation aims to optimize the overall system performance while reducing energy consumption, ensuring the sustainable operation of the system. Furthermore, balancing computational resources also involves algorithm optimization, hardware upgrades, and innovative design of network architecture.

\subsection{Data Security and Ethics}
In the underwater Metaverse, ensuring the protection of sensitive data from cyber threats while safeguarding the privacy and rights of individuals and entities will be an ongoing challenge. Due to the uniqueness of the underwater environment, traditional methods of data encryption and security protocols may no longer be applicable. Therefore, developing secure data-sharing algorithms and protocols will be the key to ensuring the security of underwater data. Furthermore, considering the potential impact of underwater activities on marine ecosystems, establishing an ethical framework\cite{Jobin2019} for the collection and use of underwater data becomes particularly important. This includes taking into account the potential effects on marine life, ensuring compliance with international environmental laws and guidelines, and maintaining the integrity of the marine environment in scientific research, exploration, and commercial activities. In this way, the underwater Metaverse can protect the marine environment and respect the rights and interests of all relevant stakeholders while innovating and advancing technology.

\section{Conclusion}
\label{conclusion}
In this article, we presented an underwater Metaverse framework, focusing not only on technology and security but also  systematic guidance and operational solutions for implementing the underwater Metaverse. We described four key technologies for the underwater Metaverse construction. For each key technology, we outlined the fundamental concepts and methods that could be used to address related issues. Then, a case study of underwater Metaverse was conducted using the proposed QI-GAI technology. The QI technology achieved high-precision and high-fidelity object capture in severe underwater environments, effectively overcoming the limitations of existing underwater imaging technologies. Generative AI technology was used to simulate and generate highly realistic underwater targets. This casew study not only demonstrated the feasibility of our proposed concepts and technologies, but also provided solid technical support for future applications in the underwater metaverse. This is the first universal framework proposed to combine underwater characteristics. Furthermore, the integration of underwater quantum imaging and GAI technology for digital content generation is an innovative experimental attempt. Finally, in future research, we will further investigate specific technical issues in the process of perception with GAI, moving towards an intelligent underwater Metaverse.

\end{document}